\documentclass{iaus}
\usepackage{graphicx}

\title[Solar butterfly diagram of 1825--1867] 
{First solar butterfly diagram from\\ 
Schwabe's observations in 1825--1867}

\author[Rainer Arlt \& Anastasia Abdolvand]   
{Rainer Arlt$^1$
 \and Anastasia Abdolvand$^2$}

\affiliation{$^1$Astrophysikalisches Institut Potsdam, \\ An der Sternwarte 16,
D-14482 Potsdam, Germany \\ email: {\tt rarlt@aip.de} \\[\affilskip]
$^2$Lyc\'e{}e Fran\c{c}ais de Berlin\\ Berlin, Germany \\
email: {\tt abdolvand\_a@hotmail.de}}

\pubyear{2010}
\volume{273}  
\pagerange{1--3}
\jname{Physics of Sun and Star Spots}
\editors{D.P. Choudhary \& K.G. Strassmeier, eds.}
\begin{document}

\maketitle

\begin{abstract}
The original sunspot observations by Heinrich Samuel Schwabe
of 1825--1867 were digitized and a first subset of spots was 
measured. In this initial project, we determined more than 14\,000 sunspot 
positions and areas comprising about 11\% of the total amount of spots 
available from that period. The resulting butterfly diagram
has a typical appearance, but with evident north-south
asymmetries.
\end{abstract}

\firstsection 
\section{Introduction}
A continuous set of sunspot positions was constructed from the 
observations at the Royal Greenwich Observatory starting in 
1874 and the observations obtained by the USAF/NOAA starting 
in 1976 (cf. Hathaway et al. 2003). Numerous investigations
regarding the butterfly diagram or sunspot area are based on this
dataset. It is desirable to extend this information back in time to
cover a larger part of the period for which sunspot {\em numbers\/} are known.
Sp\"orer (1874, 1878, 1880, 1886, 1894) gives sunspot positions for the 
period 1861--1894. A short set of observations is available from
Richard Carrington (1863) covering the period 1853--1861. An even earlier
set of sunspot positions was derived by Arlt (2009) from the observations
by Staudacher in 1749--1799.

In this paper, we present a first set of sunspot positions and areas from 
the full-disk drawings of the sun by Heinrich Samuel Schwabe at his location 
in Dessau, Germany, in 1825--1867. The original observations are preserved 
by the Royal Astronomical Society, London, and have been kindly provided
for digitization in 2009.

\section{Determination of positions and areas}
The total number of full-disk drawings
in these records is 8468 consisting of circles of 5~cm in diameter.
Within this set, 7299~drawings have a coordinate system which is
found to be aligned with the celestial equator. Especially from mid-1830
onward, almost all drawings have this coordinate system. Special
care was apparently taken by Schwabe, that the drawing represents
the situation at $12^{\rm h}$ local time. Schwabe gave descriptions
of the spots several times a day, and in nearly all cases only the 
$12^{\rm h}$ description matches the drawing. All observations were made 
looking through a Keplerian telescope equipped with one of a variety of
solar filters. All images have thus been turned by $180^\circ$ before
being used for measurements.

We assume that the middle horizontal line of the drawings is
parallel to the celestial equator and add the inclinations of
the ecliptic as well as the tilt of the solar rotation axis
against the ecliptic to the direction to the celestial pole.
A preliminary heliographic coordinate system is drawn onto
the sunspot drawing. The disks are enlarged to a size of 490~pixels
radius to achieve a slight sub-pixel accuracy as compared
to the original images. This corresponds to a scale of 
0.05 mm/pixel which is below the probable plotting errors of 
Schwabe. If that was 1~mm (which is likely to be an upper limit),
the error in heliographic position is $2.5^\circ/\cos d$, where
$d$ is the distance from the disk center. The additional
uncertainty in determining the position angle of the solar 
equator is included here with an approximate error of $0.3^\circ$; 
it is fairly well defined.

The cases without any coordinate system required special care.
We noticed that the drawings were fairly well aligned with the
celestial equator, simply by looking at the plausibility of the
spot distribution from day to day. This alignment is remarkably
consistent throughout the dataset, especially for the drawings 
explicitly marked to be made at $12^{\rm h}$ local time. We 
utilized those by assuming that an imaginary horizontal line through 
the drawing is parallel to the celestial equator. Even if one of 
Schwabe's early telescopes was not mounted parallactically, an 
alignment with the horizon in an alt--azimuth system would deliver 
the same result for observations near noon.

For a few observations before 1830, we used a special matching
algorithm for pairs of observations separated by no more than a few days.
Two or more spots need to be on both drawings. Using the differential
rotation derived by Balthasar et al. (1986), the individual 
position angles of the two drawings can be determined by a 
least-squares search. While the method may deliver additional
positions, not measurable otherwise, we need to bear in mind 
that using the differential rotation as an input reduces the
chances of an independent determination of the differential
rotation of that Cycle~7.

The sunspot size was estimated by various cursor masks with radii
from 1~to 11~pixels. As soon as Schwabe distinguished penumbra
and umbra in his drawings, we used the umbral area for the size
determinations, since we believe these are a more direct
indication for the emerging magnetic flux than the total size
of a sunspot. Note that this differs from the USAF/NOAA set
which only reports the penumbral area after Dec~16, 1981. We
also need to emphasize that we only employed the full-disk
drawings by Schwabe. Numerous detailed drawings of individual
groups are also available, showing more spots at greater
detail, but placing them to the right scale and with correct 
orientation into the full-disk drawings is a delicate task. 

\begin{figure}
\begin{center}
 \includegraphics[width=13.7cm]{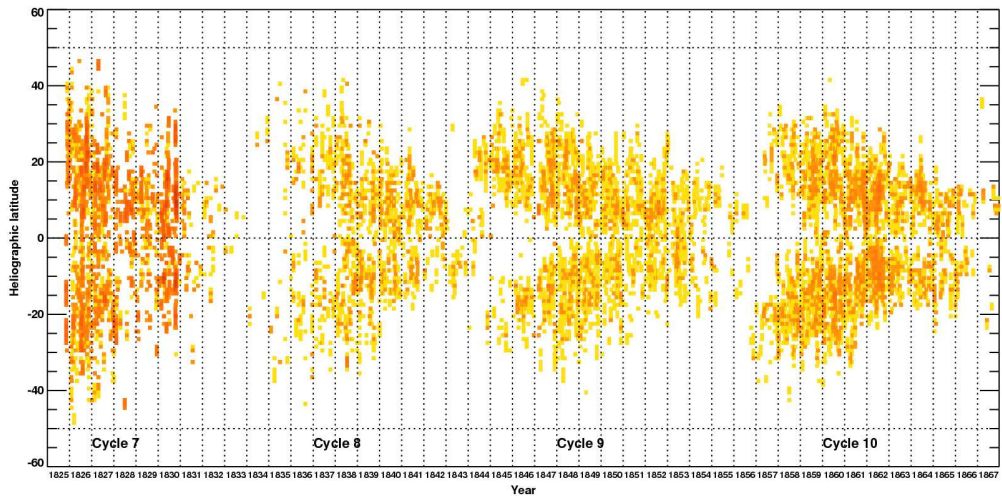} 
 \caption{Butterfly diagram obtained from the observations by Heinrich
 Schwabe in 1825--1867. The sunspot areas up to 1830 are overestimated
 since Schwabe did not distinguish penumbrae from umbrae in the beginning.}
   \label{butterfly}
\end{center}
\end{figure}

\section{Result}
The tentative butterfly diagram of the period of 1825--1867 is shown in
Fig.\,\ref{butterfly}. The darkness of the individual spots is scaled
with the area of the spots while the vertical (latitudinal) extent is 
scaled with the diameter of the spots. Note that the diagram contains only about
11\% of the full set of observations available. The accuracy will improve
significantly after measuring the remaining 89\% of the drawings.

The solar activity cycle is clearly visible. The
first cycle (Cycle~7) appears in darker colours, because Schwabe did 
not distinguish the umbra from the penumbra until 1830. The larger 
areas measured result in a ``darker'' cycle as compared to the following 
cycles. A recalibration of these data will be necessary for the final 
construction of the butterfly diagram in the future. Also, there are
some periods during Cycle~7 when we made measurements for each drawing,
while other periods were covered by every 10th drawing to get a first
picture of the entire observing period. These ``daily'' measurements
are reflected as denser stripes in the butterfly diagram. 
 
According to our results, Cycle~8 was a rather weak one, 
while the Wolf numbers indicate
a stronger cycle. Cycle~9 shows a phase difference between the
northern and southern hemispheres with the beginning of the northern 
cycle preceding the one of the southern hemisphere by at least
half a year. The analysis as in Zolotova et al. (2010), once
the full dataset is available, will show the entire development
of the phase difference for the middle of the 19th century. Cycle~10
appears to be more symmetric, and the beginning of Cycle~11 is just 
barely visible in 1867.

The data will be available from the author once the full set of
observations will be measured. 

\begin{acknowledgments}
The authors are much obliged to the Royal Astronomical Society, London,
and to Robert Massey and Peter Hingley in particular for the support 
with digitizing the original observations. We are grateful to Stela 
Frencheva, Jennifer Koch, and Christian Schmiel for their help with the 
utilization of the digital images.
\end{acknowledgments}

\end{document}